\documentclass[amssymb,amsmath,aps,showpacs,floatfix,nofootinbib,showpacs,12pt]{revtex4}
\usepackage{epsfig}
\usepackage{color}
\usepackage{graphicx}

\def\beq{\begin{equation}}
\def\eeq{\end{equation}}

\def\ev{\,{\rm eV}}
\def\be{\begin{equation}}
\def\ee{\end{equation}}
\def\bea{\begin{eqnarray}}
\def\eea{\end{eqnarray}}

\newcommand{\gsim}{\lower.7ex\hbox{$\;\stackrel{\textstyle>}{\sim}\;$}}
\newcommand{\lsim}{\lower.7ex\hbox{$\;\stackrel{\textstyle<}{\sim}\;$}}

\begin{document}
\title{Symmetry, dark matter and LHC phenomenology\\ of the minimal $\nu$SM
}

\author{ Xiao-Gang He$^{2,3}$, Tong Li$^3$, Wei Liao$^{1,3}$}
\affiliation{ $^1$ Institute of Modern Physics, East China
University of Science
and Technology, Shanghai\\
$^2$Department of Physics and Center for Theoretical
Sciences, National Taiwan University,
Taipei\\
$^3$Center for High Energy Physics, Peking University, Beijing
}

\begin{abstract}

A sterile neutrino with a mass of a few keV can play the role of a
warm dark matter(DM). This can be realized in seesaw models with 3
left- and 3 right-handed neutrinos. It is possible to identify the
keV neutrino to be one of the right-handed neutrinos leaving the
other two to be much more heavier, the $\nu$SM model. We show that
with this realization of keV neutrino DM, the model has an
approximate Friedberg-Lee symmetry providing a natural explanation
for the lightness of the right-handed neutrino. We also find that in
this model the mixing parameters couple light and heavy neutrinos
are strongly correlated, and can be large enough to have testable
effects at the LHC for the two heavy right-handed neutrinos to be in
the hundred-GeV range.

\end{abstract}


\maketitle

 \noindent {\bf Introduction}

Cosmological and astrophysical studies show that there are dark
matter (DM) in our universe. The DM contributes about 20\% of the
energy in our universe. The identity of DM is still not known. Many
models have been proposed. Neutrino has long been considered to be
one of the possible candidates. The left-handed neutrinos, the
active neutrinos, have standard model (SM) weak interaction, and
were in thermal equilibrium in the early universe. Active neutrinos
with masses in the range of a few tens of eV to a few GeV would over
close the universe and are therefore ruled out as DM. Active light
neutrinos of mass less than a few tens of eV has problem with
structure formation. Data constrain the sum of the three light
masses must be less than an eV or so. An active neutrino is unlikely
to play a significant role for DM. However, a right-handed neutrino
$\nu_R$ with an appropriate mass and a mixing with active neutrinos
can play the role of DM. The right-handed neutrino does not have SM
interactions and may not be in thermal equilibrium in the early
universe. But in general they mix with left-handed neutrinos and may
be produced in the early universe through oscillations of different types of neutrinos. With
appropriate mass and mixing, a right-handed neutrino of mass a few
keV can contribute the correct relic density of our universe. This
$\nu_{R}$ DM belongs to the warm DM category.

Right-handed neutrinos can be introduced in different ways. If right
handed neutrinos have large Majorana masses, they can play a very
important role to explain why the light neutrinos have very small
masses via the seesaw mechanism~\cite{seesaw1}. A priori, the
right-handed neutrino mass scale is not known which can be as high
as the Planck scale or as low as 1 eV leading to many interesting
consequences~\cite{seesaw1,ABS,AS,he,gouvea}. We will study some
implications of a seesaw model with a keV mass right-handed neutrino
playing the role of DM companied by two heavy neutrinos. A model,
called $\nu$MSM to realize this has been proposed in~\cite{ABS,AS}.
It is a seesaw model~\cite{seesaw1} in which there are 3 left- and 3
right- handed neutrinos, the 3+3 model, with one of the right-handed
neutrinos ($\nu_{R_1})$ having a mass of a few keV. This is a
minimal model of this type. In the model proposed in Refs.
~\cite{ABS,AS} the heavy neutrinos $\nu_{R2,3}$ have masses around
$1-10$ GeV. In the present work we take a different approach to have
$\nu_{R_{2,3}}$ mass scale to be in the hundred-GeV range and to
study some implications for LHC physics. We refer this model as
$\nu$SM.

In order for the keV right-handed neutrino to play the role of a
warm DM, the parameters in this model are constrained. We show that
the model has an approximate Friedberg-Lee symmetry providing an
natural explanation for the lightness of the keV right-handed
neutrino. We also find that in this model the mixing parameters
which couple light and heavy neutrinos are strongly correlated, and
can be large enough to have testable consequences at the LHC.

\vspace{0.3cm}
 \noindent {\bf Dark matter in $\nu$SM}

Now we briefly describe the warm DM in $\nu$SM. Let us indicate the
keV scale right-handed neutrino which plays the role of the DM as
$\nu_{R_1}$, and the other two right-handed neutrinos which have
masses in the multi-GeV or higher region as $\nu_{R_{2,3}}$.
The other particles relevant to our discussions
are the three generations of left-handed lepton doublets
 $L_{Li} = (\nu_{Li}, e_{Li})^T$, and the Higgs doublet $H = (H^0, H^-)^T$.
 The Lagrangian responsible to neutrino masses is
 \begin{eqnarray}
{\cal L} =  - {1\over 2}
\bar \nu_R M \nu^c_R  - \bar L_L Y H \nu_R + H.C. \;,\label{final-form}
\end{eqnarray}
 where $\nu^c_{R}$ is the charge conjugate of $\nu_R$. $M$ and $Y$ are $3\times 3$ matrices. $M$ is the Majorana mass
 matrix of $\nu_R$ and is symmetric. For convenience we will work in the basis where $M$ is
 diagonal, that is, $M=diag \{M_1,M_2,M_3\}$.

 In seesaw models one usually assumes that the
 right-handed neutrinos are super-heavy in the $10^{14} \sim 10^{15}$ GeV range, so
 that the masses of light left-handed neutrinos are strongly
 suppressed. But the scale need not to be so high. In fact the scale can be as low as a few hundreds of GeV to a TeV.
 In Ref. \cite{ABS} it was suggested that the lightest right-handed neutrino $\nu_{R_1}$ can even have a mass a few keV and play the role of a warm
 DM.

After the electro-weak symmetry breaking, that is, the Higgs develops
a non-zero vacuum expectation value $\langle H\rangle = (v,0)^T$,
the neutrino mass matrix
 in the basis $(\nu_L, \nu_R^c)^T$ is given by
 \begin{eqnarray}
 \left (\begin{array}{cc} 0&Y^*v \\Y^\dagger v &M \end{array} \right )\;.
 \end{eqnarray}
 Diagonalization of the above seesaw mass matrix leads to the reduced mass matrix $m_\nu$ for
 light neutrinos,
 \begin{eqnarray}
 m_\nu = -v^2 Y^* M^{-1} Y^\dagger. \label{Numass}
 \end{eqnarray}
 The above seesaw mass matrix also leads to a non-zero mixing matrix between the
 left- and right- handed neutrinos given by
 \begin{eqnarray}
 R = (R_{li}) =Y v (M^*)^{-1}, \label{mixing1}
 \end{eqnarray}
 where $l$ runs over the light neutrinos, $\nu_{e,\mu,\tau}$ and $i$ runs over, $\nu_{R_{1,2,3}}$.

 For keV scale $\nu_{R_1}$, the above mixing can cause
oscillation, in the early universe, between the right- and left-
handed neutrinos and lead to a finite energy density from the keV
right-handed neutrino~\cite{DW,Abazajian,ABS},
\begin{eqnarray}
\Omega_{\nu_R} h^2 \sim 0.1 ~{\sum_l |R_{l1}|^2\over 10^{-8}}\left
({M_1\over 3 ~\textrm{keV} }\right )^2\;. \label{relic}
\end{eqnarray}
Eq. (\ref{relic}) is for production without resonance. For
production with resonance, sufficient amount of keV scale $\nu_{R1}$
warm DM can be produced for a much smaller mixing~\cite{blrv}.

It is interesting that $\nu_{R_1}$ of a few keV can provide the
right amount of warm DM density if it is stable enough. The keV
scale $\nu_{R_1}$ can only decay into light active
 neutrinos at the tree-level through mixing of left- and right-handed neutrinos and
 exchanging $Z$ and Higgs. At loop level it can also decay into photon
 through exchanging W and charged leptons. The tree-level decay modes of
 $\nu_{R_1}$ are: $\nu_{R_1}\to \nu+2 {\bar \nu}$ and $\nu_{R_1}\to 2 \nu+ {\bar
 \nu}$. The lifetime of $\nu_{R_1}$ is estimated to be~\cite{ABS}
 \begin{eqnarray}
 \tau_{\nu_{R1}}=5. \times 10^{26} \textrm{s}
 ~\bigg(\frac{1 ~\textrm{keV}}{M_1} \bigg)^5 \frac{10^{-8}}{\sum_l|R_{l1}|^2}.
 \end{eqnarray}
 One can see that the lifetime of $\nu_{R_1}$ is much larger than the
 age of the universe $ \sim 10^{17}$s for mass $M_{R1}$ of order a keV
 and  $\sum_l|R_{l1}|^2\sim 10^{-8}$. A keV scale $\nu_{R_1}$ has a long lifetime
 which is allowed for a DM candidate.

 Constraints on the mass and the mixing of $\nu_{R_1}$ as warm DM
 candidate come from structure formation~\cite{ABS}, phase space density~\cite{bri},
 X-ray lines of $\nu_{R_1} \to \nu_a({\bar \nu}_a) +\gamma$ decay~\cite{bnrst,bnr,birs} and
 Lyman-$\alpha$ forest data~\cite{blrv2}. Astrophysical implications
 of keV scale $\nu_R$, such as effect on supernova explosion and
 re-ionization etc., have been explored in Refs.
 ~\cite{supernova,reionization}.
 A review on related subjects can be found in Ref.~\cite{Kusenko:2009up}.
 Considerations on
 structure formation, phase space density and Lyman-$\alpha$ forest data
 give lower bound on the mass of $\nu_{R_1}$ warm DM. Observations
 of X-ray lines of $\nu_{R_1}$ decay give upper bound on the
 mass and mixing of $\nu_{R_1}$. These constraints can be very strong
 if $\nu_{R_1}$ warm DM accounts for all the DM density in the
 universe. However, a recent analysis ~\cite{blrv2} shows that these constraints
 can be significantly released for $\Omega_{\nu_R} \lsim 0.4 ~\Omega_{DM}$.
 For example,  $\nu_{R_1}$ with mass in the range
 $1$ keV $\lsim M_1 \lsim 5$ keV and $\sum_l|R_{l1}|^2\sim 10^{-9}-10^{-8}$
 is allowed by all these constraints if $\Omega_{\nu_{R1}} = 0.2~ \Omega_{DM}$.
 We will use $1$ keV $\lsim M_1 \lsim 5$ keV and
 $|R_{l1}|^2 \lsim 10^{-8}$ for discussions later.

 The other two right-handed neutrinos, $\nu_{R_{2,3}}$, can have large masses and
 generate the light active neutrino
 masses and mixing through the seesaw mechanism.
 This model can correctly account for the masses and mixing
 of light neutrinos by making appropriate choice of Yukawa couplings~\cite{ABS}.

\vspace{0.3cm}
 \noindent {\bf Implication of the keV neutrino for the Yukawa couplings}

 In general the Yukawa coupling matrix  $Y$ can be parameterized in the following form~\cite{ci}
 \begin{eqnarray}
  Y = {1\over v}U ({\tilde m}^*_\nu)^{1/2} O (M^*)^{1/2} , \label{see-saw3}
 \end{eqnarray}
 where ${\tilde m}^{1/2}_\nu
=\textrm{diag} \{ m_1^{1/2} e^{i\phi_1/2},
 m_2^{1/2} e^{i\phi_2/2}, m_3^{1/2} e^{i\phi_3/2} \}$
 where real numbers $m_i(i=1,2,3)$ are the masses of three light
 neutrinos and $\phi_i(i=1,2,3)$ are three Majorana phases.
 $(M^*)^{1/2} =\textrm{diag}\{(M^*_1)^{1/2}, (M^*_2)^{1/2}, (M^*_3)^{1/2}\}$.
 $U$ is the neutrino mixing matrix observed in experiments,
 ${\tilde m}_\nu$ is the diagonalized neutrino mass matrix, $O$ is
 a complex orthogonal matrix: $O^T O=O O^T=1$.

 Using Eq. (\ref{see-saw3}), the mixing matrix $R$ can be written as
 \begin{eqnarray}
 R=U ({\tilde m}^*_\nu)^{1/2} O (M^*)^{-1/2}. \label{mixing3}
 \end{eqnarray}

 The orthogonal matrix $O$ can have determinant $det(O)=\pm 1$.
 Orthogonal matrix $O$ with det$(O)=+ 1$ can be parameterized as the products
 of three matrices:
 \begin{eqnarray}
 O=O_{23} O_{13} O_{12}, \label{Omatrix1}
 \end{eqnarray}
 where $O_{ij}$ is a $3\times 3$ complex orthogonal matrix with determinant
 $+1$. It can be expressed using complex numbers $\theta_{ij}$, e.g.
 \begin{eqnarray}
 O_{12} =
 \begin{pmatrix}
                          \cos\theta_{12} & \sin\theta_{12} & 0\cr
                         -\sin\theta_{12} & \cos\theta_{12} & 0 \cr
                          0 & 0 & 1
 \end{pmatrix}.
 \label{Omatrix2}
 \end{eqnarray}

 Orthogonal matrix $O$ with det$(O)=-1$ differs from
 an orthogonal matrix expressed in Eq. (\ref{Omatrix1})
 by a transformation using matrices such as diag$\{1,1,-1\}$, diag$\{1,-1,1\}$, etc.
 These factor $-1$ can be absorbed into the phase factors in $({\tilde
 m}^*_\nu)^{1/2}$. That is, we can choose
 \begin{eqnarray}
 0 \leq \phi_i < 4 \pi \label{phaserange}
 \end{eqnarray}
 instead of in the range $[0,2 \pi]$ to account for the possibility
 that det$(O)$ can be $-1$.

 To further understand implications of $\nu$SM we rewrite the seesaw mass
 formula, Eq. (\ref{Numass}), as the following
 \begin{eqnarray}
 m_\nu =\sum_i S_i, ~~i=1,2,3   \label{mass2}
 \end{eqnarray}
 where
 \begin{eqnarray}
 (S_i)_{ll'}=-v^2 Y^*_{li} Y^*_{l'i} M^{-1}_i=- M_i R^*_{li} R^*_{l'i}. \label{mass2b}
 \end{eqnarray}
 Eq. (\ref{mixing1}) has been used in obtaining Eq. (\ref{mass2b}).
 $S_i$ is the contribution of $\nu_{Ri}$ to the light neutrino masses.
 Using the constraint $\sum_l |R_{l1}|^2 \lsim 10^{-8}$, we find for
 $1$ keV $ \lsim M_1 \lsim 5$ keV~\cite{ABS}
 \begin{eqnarray}
 |(S_1)_{ll'}| \lsim 10^{-5} ~\textrm{eV}. \label{constraint1}
 \end{eqnarray}

 We introduce
 \begin{eqnarray}
 Y' && \equiv U^\dagger Y={\tilde m}^{1/2}_\nu O (M^*)^{1/2} v^{-1},
 \label{newbase1} \nonumber\\
 R' && \equiv U^\dagger R={\tilde m}^{1/2}_\nu O (M^*)^{-1/2},
 \label{newbase2}\nonumber \\
 (S'_i)_{ab} && \equiv U^T S_i U=-v^2 Y^{'*}_{ai} Y^{'*}_{bi}
 M_i^{-1}.
 \label{newbase3}
 \end{eqnarray}
 Using Eq. (\ref{newbase1}) we get
 \begin{eqnarray}
 (S'_i)_{ab} =- m_a^{1/2} m_b^{1/2} e^{i(\phi_a+\phi_b)/2}
 O^*_{ai} O^*_{bi}.
 \label{newbase4}
 \end{eqnarray}
 The condition Eq. (\ref{constraint1}) is re-expressed as
 \begin{eqnarray}
 |(S'_1)_{ab}| \lsim 10^{-5} ~\textrm{eV} \label{constraint1a}
 \end{eqnarray}

The experimental constraints on the neutrino masses, at $2\sigma$
level~\cite{Schwetz:2008er}, are
 \bea
7.25 \times 10^{-5} \ev^2 \  < & \Delta m_{21}^2 & < \  8.11 \times 10^{-5} \ev^2, \\
2.18 \times 10^{-3} \ev^2 \  < & |\Delta m_{31}^2| & < \  2.64
\times 10^{-3} \ev^2,
 \label{Mass}
  \eea
  $\sum_{i} m_{i} < \ 1.2 \ \ev$.
 For normal hierarchy(NH) of neutrino masses,
 $m_2\approx \sqrt{\Delta m^2_{21}}\approx 0.9 \times 10^{-2}$ eV
 and $m_3\approx \sqrt{|\Delta m^2_{31}|}\approx 0.05$ eV,
 Eq.~(\ref{constraint1a}) says $|(S'_1)_{aa}|=m_a |O^{*2}_{a1}| \ll m_a$
 for $a=2,3$. That is
 \begin{eqnarray}
 |O_{a1}| \ll 1, ~~a=2,3 \label{approxNH}
 \end{eqnarray}
 From this condition we find that the matrix $O$ is approximated as
 \begin{eqnarray}
 O\approx \begin{pmatrix} 1 & 0 & 0 \cr
                          0 & \cos \theta & \sin\theta \cr
                          0 & -\sin \theta & \cos\theta
 \end{pmatrix}.
 \label{NHmatrix}
 \end{eqnarray}

 For inverted hierarchy(IH) of neutrino masses,
 $m_1\approx \sqrt{|\Delta m^2_{31}|}\approx 0.05$ eV
 and  $m_2\approx \sqrt{|\Delta m^2_{31}|+\Delta m^2_{21}}\approx 0.05$ eV,
 (\ref{constraint1a}) says $|(S'_1)_{aa}|=m_a |O^{*2}_{a1}| \ll m_a$
 for $a=1,2$. That is
 \begin{eqnarray}
 |O_{a1}| \ll 1, ~~a=1,2
 \end{eqnarray}
 From this condition we find that the matrix $O$ is approximated as
 \begin{eqnarray}
 O\approx \begin{pmatrix} 0 & \cos\theta & \sin\theta \cr
                          0 & -\sin \theta & \cos\theta \cr
                          1 & 0 & 0
 \end{pmatrix}.
 \label{IHmatrix}
 \end{eqnarray}

 The above analysis show that $\nu_{R_1}$ gives negligible contribution to light neutrino masses.
 This condition implies that light neutrino masses are dominated by contributions
 of two heavier right-handed neutrinos $\nu_{R2,3}$. The lightest left-handed
 neutrino has mass of order $\sim 10^{-5}$ eV. $\nu$SM can not reproduce degenerate
 mass pattern of light left-handed neutrinos. We also have a good idea about
 some of the properties of the Yukawa coupling matrix $Y$. We will show in the following that the resulting
 mass matrix exhibits an approximate Friedberg-Lee(FL) symmetry~\cite{Friedberg:2006it, Friedberg:2007ba}.

\vspace{0.3cm}
 \noindent {\bf The approximate Friedberg-Lee symmetry in $\nu$SM }

 A theory is said to have a FL symmetry
 when the Lagrangian of this theory is invariant under the transformation on a fermion field of the form
 $q \to q + \epsilon$, where $\epsilon$ is a space-time independent element of the Grassmann
 algebra, anti-commuting with the fermionic field operators $q$. We explain why the neutrino mass matrix in the $\nu$SM has an
 approximate FL symmetry in detail in the following.

The general Yukawa matrix can be written as
 \begin{eqnarray}
  Y= \begin{pmatrix} {\hat Y}_{e1} & {\tilde Y}_{e 2} & {\tilde Y}_{e 3} \cr
 {\hat Y}_{\mu 1} & {\tilde Y}_{\mu 2} & {\tilde Y}_{\mu 3} \cr
 {\hat Y}_{\tau 1} & {\tilde Y}_{\tau 2} & {\tilde Y}_{\tau 3} \cr
 \end{pmatrix}. \label{Yb}
 \end{eqnarray}

 Using Eqs. (\ref{NHmatrix}) and (\ref{IHmatrix}) we find that for NH case, for ${\tilde Y}$ part of the $Y$ one has
 \begin{eqnarray}
 v {\tilde Y}= U ({\tilde m}^*_\nu)^{1/2}
 \begin{pmatrix} 0 & 0 \cr
                 \cos \theta & \sin\theta \cr
                 -\sin \theta & \cos\theta
 \end{pmatrix}  ({\tilde M}^*)^{1/2}, \label{NHmatrix1}
 \end{eqnarray}
 where ${\tilde M}=diag \{M_2,M_3 \}$.

 For IH case ${\tilde Y}$, one has
 \begin{eqnarray}
 v {\tilde Y}= U ({\tilde m}^*_\nu)^{1/2}
 \begin{pmatrix}
                 \cos \theta & \sin\theta \cr
                 -\sin \theta & \cos\theta \cr
                     0 & 0
 \end{pmatrix}  ({\tilde M}^*)^{1/2}. \label{IHmatrix1}
 \end{eqnarray}
 while the ${\hat Y}$ part of $Y$ is given by
 \begin{eqnarray}
 v {\hat Y}= U ({\tilde m}^*_\nu)^{1/2}
 \begin{pmatrix} (M^*_1)^{1/2} \cr
                 0 \cr
                     0
 \end{pmatrix} ~\textrm{or}~ v {\hat Y}=U ({\tilde m}^*_\nu)^{1/2}
 \begin{pmatrix} 0 \cr
                 0\cr
                (M^*_1)^{1/2} \label{Yhat}
 \end{pmatrix},
 \end{eqnarray}
 for NH or IH, respectively.

 Using result from Eq. (\ref{constraint1a}) we can also write
  the light neutrino mass matrix as
 \begin{eqnarray}
 m_\nu \approx - v^2 {\tilde Y}^* {\tilde M}^{-1} {\tilde
 Y}^\dagger. \label{minimal-seesaw}
 \end{eqnarray}
 The above is the mass formula in the minimal seesaw
 model~\cite{Frampton:2002qc}.

 Note that since $|M_{2,3}|\gg |M_1|$ and the ${\hat Y}_{l1}$ is
 suppressed by $M^{1/2}_{1}/M_{2,3}^{1/2}$ in comparison with
 elements of ${\tilde Y}$.
 In Eqs. (\ref{NHmatrix}) and (\ref{IHmatrix}) there are no large
 factors in the first columns to enhance the Yukawa couplings,
 ${\tilde Y}$ is the leading term in Yukawa coupling, we have approximately:
 \begin{eqnarray}
 Y\approx \begin{pmatrix} 0 & {\tilde Y}_{e 2} & {\tilde Y}_{e 3} \cr
 0 & {\tilde Y}_{\mu 2} & {\tilde Y}_{\mu 3} \cr
 0 & {\tilde Y}_{\tau 2} & {\tilde Y}_{\tau 3} \cr
 \end{pmatrix}. \label{Yapprox}
 \end{eqnarray}
  The mass matrix $M$ can be
 approximated as
 \begin{eqnarray}
 M \approx \textrm{diag} \{0, M_2, M_3 \}. \label{Mapprox}
 \end{eqnarray}

 It is easy to check that with the above $Y$ and $M$, the Lagrangian in
 Eq.(\ref{final-form}) is invariant under a FL transformation
 \begin{eqnarray}
 \nu_{R_1}\to \nu_{R_1}+\epsilon \label{FL}
 \end{eqnarray}
 If the FL symmetry is a global one, that is, $\epsilon$ is space-time independent,
 it is easy to check that the kinetic term
 ${\cal L}_k = \bar \nu_R \gamma_\mu (i \partial^\mu \nu_R)$ is also invariant under
 the transformation defined in Eq.(\ref{FL}).

 In fact it has been shown~\cite{HeLiao,Jarlskog:2007qy} that imposing a global FL symmetry on a
 particular direction of right-handed neutrinos in the 3+3,
 it results in a massless right-handed neutrino, and the mass matrices is
 equivalent to the from of a 3+2 (two right-handed neutrinos) minimal seesaw model
 which predicts a light neutrino with a zero mass.  In our case, $\nu_{R_1}$ is
 the corresponding massless right-handed neutrino, and the lightest light neutrino
 mass of order $10^{-5}$ eV corresponds to the zero mass one in the exact FL limit.

 We conclude that the $\nu$SM has an approximate FL symmetry. Small violation of
 this symmetry provides a natural explanation why one of the right-handed neutrino
 has much smaller mass than other two heavy right-handed neutrinos. We would
 like to comment that the 3+3 seesaw model is the minimal model which is consistent
 with light neutrino masses and can have a warm DM candidate in the framework of
 seesaw mechanism. We note that the FL symmetry we found is a consequence of the
 experimental constraints. Some other symmetries can be imposed to obtain
 $\nu$SM~\cite{Shap}.

 \vspace{0.3cm}
 \noindent {\bf The possibility of a large mixing between light and heavy sectors in $\nu$SM }

 The model considered in Ref.~\cite{ABS} has very small Yukawa
 couplings. They assume the Yukawa couplings are of order
 $\sim m_\nu^{1/2} M^{1/2}_j/v$. If this is always the case, then even the heavy right-handed neutrinos are light enough,
 a few hundred GeV, to be produced at the LHC, the small mixing between
 light and heavy neutrinos makes it impossible to be detected.
 Fortunately this is not necessarily true. There are other possibilities\cite{he}.

 It is known that the matrix elements of the complex orthogonal
 matrix $O$ can be large if $\theta$ is a complex number:
 \begin{eqnarray}
 \theta=x+iy
 \end{eqnarray}
 where $x$ and $y$ are two real numbers. In this case $|\cos\theta|$
 and $|\sin\theta|$ can be enhanced by a large factor $e^{|y|}$ if $|y|$ is large.
 Elements in the Yukawa coupling $Y$ matrix and the mixing
 $R$ matrix can be enhanced by large elements of matrix $O$. Tiny light neutrino
 masses are reproduced with large Yukawa couplings in the seesaw formula
 through fine tuning. In the following we explain
 in more detail how a large mixing between light and heavy neutrinos can be obtained in $\nu$SM.

 Consider the NH case. Eq. (\ref{NHmatrix}) can be written as
 \begin{eqnarray}
 O\approx \frac{1}{2} e^{\mp ix+|y|}\begin{pmatrix} 0 & 0 & 0 \cr
                          0 & 1 & \pm i \cr
                          0 & \mp i & 1
 \end{pmatrix} + \begin{pmatrix} 1 & 0 & 0 \cr
                          0 & \frac{1}{2} e^{\pm ix-|y|}& \mp \frac{i}{2} e^{\pm ix-|y|}\cr
                          0 & \pm \frac{i}{2}e^{\pm ix-|y|} & \frac{1}{2}
                          e^{\pm ix-|y|}
 \end{pmatrix}, ~\textrm{for} ~y=\pm |y|.
 \label{NHmatrix2}
 \end{eqnarray}
 For $|y|\gg 1$,  the first term in the right-handed side of Eq. (\ref{NHmatrix2})
 is enhanced by the large factor $e^{|y|}$ and is the leading term.

 Using Eqs. (\ref{see-saw3}) and (\ref{NHmatrix2}) we find for $|y|\gg 1$
 the leading term in Yukawa coupling is
 \begin{eqnarray}
 v Y_{l2} && =\frac{1}{2} e^{\mp i x+|y|}( U_{l2} m^{1/2}_2
 e^{-i \phi_2/2}\mp i U_{l3} m^{1/2}_3 e^{-i\phi_3/2}) (M^*_2)^{1/2},
 \label{NHYukawa1} \\
 v Y_{l3} && =\frac{1}{2} e^{\mp i x+|y|}( \pm i U_{l2} m^{1/2}_2
 e^{-i \phi_2/2}+  U_{l3} m^{1/2}_3 e^{-i\phi_3/2})
(M^*_3)^{1/2}.
 \label{NHYukawa2}
 \end{eqnarray}
 $|Y_{l1}|$ are small as explained in Eq. (\ref{Yhat}).

 For IH case we find
\begin{eqnarray}
 O\approx \frac{1}{2} e^{\mp ix+|y|}\begin{pmatrix} 0 & 1 & \pm i \cr
                          0 & \mp i & 1 \cr
                          0 & 0 & 0
 \end{pmatrix} + \begin{pmatrix}
                          0 & \frac{1}{2} e^{\pm ix-|y|}& \mp \frac{i}{2} e^{\pm ix-|y|}\cr
                          0 & \pm \frac{i}{2}e^{\pm ix-|y|} & \frac{1}{2}
                          e^{\pm ix-|y|} \cr
                          1 & 0 & 0
 \end{pmatrix}, ~\textrm{for} ~y=\pm |y|.
 \label{IHmatrix2}
 \end{eqnarray}
 Again for $|y|\gg 1$,  the first term in the right-handed side of Eq. (\ref{IHmatrix2})
 is the leading term. The leading terms in the Yukawa coupling are
 found to be
\begin{eqnarray}
 v Y_{l2} && =\frac{1}{2} e^{\mp i x+|y|}( U_{l1} m^{1/2}_1
 e^{-i \phi_1/2}\mp i U_{l2} m^{1/2}_2 e^{-i\phi_2/2}) (M^*_2)^{1/2},
 \label{IHYukawa1} \\
 v Y_{l3} && =\frac{1}{2} e^{\mp i x+|y|}( \pm i U_{l1} m^{1/2}_1
 e^{-i \phi_1/2}+  U_{l2} m^{1/2}_2 e^{-i\phi_2/2})
(M^*_3)^{1/2}.
 \label{IHYukawa2}
 \end{eqnarray}

 From Eqs. (\ref{NHYukawa1}), (\ref{NHYukawa2}), (\ref{IHYukawa1})
 and (\ref{IHYukawa2}) we find that
 \begin{eqnarray}
 Y_{l3} (M^*_3)^{-1/2} \approx \pm i Y_{l2} (M^*_2)^{-1/2}.
 \label{testable-nuSM1}
 \end{eqnarray}
 Using Eq. (\ref{mixing1}), Eq. (\ref{testable-nuSM1}) is re-expressed as
 \begin{eqnarray}
 R_{l3} (M^*_3)^{1/2} \approx \pm i R_{l2} (M^*_2)^{1/2}.
 \label{testable-nuSM2}
 \end{eqnarray}
 Using Eqs. (\ref{testable-nuSM1}) and (\ref{testable-nuSM2}),
 we see in Eq. (\ref{mass2b}) that there is a strong cancelation
 between $S_2$ and $S_3$, contributions of $\nu_{R2}$ and $\nu_{R3}$
 to light neutrino masses.

 It has been shown in Ref. \cite{neutrinoless} that neutrinoless double
 decay gives strong constraint on the mixing and masse of sterile
 neutrino. Typically constraint for a single sterile neutrino is
 found to be $|R_{es}|^2\lsim 10^{-5}$ for $M_{s}\sim 100$ GeV. We note
 that this constraint does not apply to $\nu_{R2,3}$ of degenerate or
 quasi-degenerate masses. GeV scale $\nu_{R2,3}$ contribute to the
 neutrinoless double beta decay with the amplitude
 \begin{eqnarray}
 {\cal A}= F (R_{e2}^2 \frac{1}{M_2}+R_{e3}^2 \frac{1}{M_3}),
 \label{doublebeta}
 \end{eqnarray}
 where $F$ is a factor containing all other effects and $M_{2,3}$ have
 been chosen real for convenience of later
 discussion. Eq. (\ref{doublebeta}) can be rewritten as
\begin{eqnarray}
 {\cal A}= \frac{F}{M_2^2} (R_{e2}^2 M_2+R_{e3}^2 M_3)
 + F R_{e3}^2 M_3 (\frac{1}{M_3^2}-\frac{1}{M_2^2}).
 \label{doublebeta1}
 \end{eqnarray}
 According to Eq. (\ref{testable-nuSM2}) and Eq. (\ref{mass2b}),
 the first term in Eq. (\ref{doublebeta1}) has a strong cancelation and
 is of order $(F /M_2^2)(m_\nu)_{ee}$ and is negligible. The second
 term in Eq. (\ref{doublebeta1}) is essential for constraining the
 $\nu$SM. It's clear that if $\nu_{R2,3}$ are degenerate or quasi-degenerate,
 their contribution to neutrinoless double decay is greatly
 reduced. For example, if $(\Delta M^2_{32}/ M_2^2) \sim 10^{-5}$
 where $\Delta M^2_{32}=M_3^2-M_2^2$, the rate of double beta decay is
 reduced by 10 orders of magnitude and the constraint becomes  $|R_{e2,3}|^2 \lsim 1$.
 We note that a quasi-degeneracy of heavy neutrinos is required for producing
 baryogenesis using oscillation of right-handed neutrinos~\cite{AS,ARS} or
 using thermal leptogenesis~\cite{FY} with resonance at the electro-weak scale.

 Matrix $Y$ can be transformed using bi-unitary transformation to a
 diagonalized form: ${\tilde Y} =\textrm{diag} \{y_1,y_2,y_3\}$.
 Large matrix elements in Eqs. (\ref{NHmatrix2}) and
 (\ref{IHmatrix2}) affect $y_{2,3}$ which correspond to
 two linear combinations of quasi-degenerate $\nu_{R2,3}$.
 Note that the matrix $O$ does not change the determinant of $Y$.
 Hence one of $y_{2,3}$, which we take to be $y_3$, is enhanced and another
 one, $y_2$, is suppressed. The fact that only one of $y_{1,2,3}$ is enhanced
 can be seen clearly in Eqs.(\ref{NHmatrix2}) and (\ref{IHmatrix2}), that is the
 leading part of the $O$ matrix, enhanced by $e^{|y|}$, is of rank one.
 Hence one can make $y_2\lsim 10^{-7}$ and $y_3$ large. This means one
 of the linear combination of quasi-degenerate
 $\nu_{R2,3}$ can never be in thermal equilibrium before the freeze-out of the
 sphaleron transition which is necessary for producing baryogenesis.
 Whether with large elements in $O$ realistic baryongenesis can be obtained is
 an interesting question to study. Detailed study on this subject is out of the scope of the
 present article and will be studied elsewhere. Here we are interested to see if
 there are parameter spaces in which experiments at the LHC can probe.

 We summarize some of the interesting properties in the following:
 \begin{itemize}
 \item
 The Yukawa couplings $Y_{l2,l3}$ are enhanced by large $e^{|y|}$ if
 $|y|$ is large.

 \item
Only one of the Majorana phases $\phi_i$ is observable. In both
NH and IH cases
 we can set $\phi_1=\phi_3=0$ and keep $\phi_2$. For convenience we write
 \begin{eqnarray}
 ({\tilde m}_\nu^*)^{1/2}=\textrm{diag} \{m_1^{1/2}, m_2^{1/2} e^{i \Phi},
 m_3^{1/2}\},
 \end{eqnarray}
 where $\Phi=-\phi_2/2$. $\Phi$ can be chosen in the range
 $[0,2\pi]$ as discussed for Eq. (\ref{phaserange}).

\item
 None of the Majorana phases in $M_i$ is observable since one can
 always rotate away phases of $M_i$.

 \item
 A CP violating phase $\mp x$ in the factor $e^{\mp i x}$ is
 observable.

 \item
 For GeV scale degenerate $\nu_{R2,3}$ the constraint from the
 neutrinoless double beta decay on the mixing $R_{ei}$ is greatly reduced.

 \end{itemize}

The right-handed heavy neutrinos do not have SM gauge interaction.
If they do not mix with left-handed neutrinos or the mixing is
extremely small, it is not possible to produce them and study their
properties. As we have shown in the above that it is possible to
have large mixing. The production of heavy neutrinos at the LHC
becomes possible. Using constraints from electro-weak precision
test, it has been shown model-independently that the heavy neutrinos
can be produced and studied at the LHC up to 400 GeV for $2\sigma$
with 100 fb$^{-1}$ of integrated luminosity and mixing
$|R_{li}|^2\lesssim 10^{-3}$~\cite{Tao}. The mixing of $\nu_{R_1}$
with left-handed neutrinos are too small to have direct laboratory
observable effects. But $\nu_{R_{2,3}}$ may be produced and their
properties may be studied.
 We note that $|R_{ei}|^2 \gsim
 10^{-3}$ can be reached in $\nu$SM, consistent with constraint from
 double beta decay experiments, for quasi-degenerate
$\nu_{R2,3}$ which have $\Delta M^2_{32}/M_2^2 \lsim 10^{-2}$. In
the later discussion of the decay pattern of $\nu_{R2,3}$ we will
concentrate on the degenerate or quasi-degenerate case.

\vspace{0.3 cm}
\noindent {\bf Correlation of $\nu_{R_{2,3}}$ and light neutrino mass hierarchy in $\nu$SM }

As seen from Eqs. (\ref{see-saw3}) and (\ref{mixing3}), for a given
form of $O$, one can establish the connection between the Yukawa
coupling $Y$, the mixing matrix $R$, and the properties of the light
neutrinos ($U$ and masses).  The leptonic mixing matrix $U$ is usually written as
\begin{equation}
U= \left(
\begin{array}{lll}
 c_{12} c_{13} & c_{13} s_{12} & e^{-\text{i$\delta $}} s_{13}
   \\
 -c_{12} s_{13} s_{23} e^{\text{i$\delta $}}-c_{23} s_{12} &
   c_{12} c_{23}-e^{\text{i$\delta $}} s_{12} s_{13} s_{23} &
   c_{13} s_{23} \\
 s_{12} s_{23}-e^{\text{i$\delta $}} c_{12} c_{23} s_{13} &
   -c_{23} s_{12} s_{13} e^{\text{i$\delta $}}-c_{12} s_{23} &
   c_{13} c_{23}
\end{array}
\right)
\end{equation}
where $s_{ij}=\sin{\theta_{ij}}$, $c_{ij}=\cos{\theta_{ij}}$, $0 \le
\theta_{ij} \le \pi/2$ and $0 \le \delta \le 2 \pi $. The phase
$\delta$ is the Dirac CP phase. The experimental constraints on the
neutrino masses are listed in Eq. (\ref{Mass}) and the constraints
on the mixing parameters, at $2\sigma$ level~\cite{Schwetz:2008er},
are
 \bea
                   0.27 \  < & \sin^2{\theta_{12}} & < \  0.35, \nonumber\\
                   0.39 \  < & \sin^2{\theta_{23}} & <\  0.63, \\
                          & \sin^2{\theta_{13}} & <\  0.040.\nonumber
                          \label{Mixing}
\eea

\begin{figure}[tb]
\begin{center}
\begin{tabular}{cc}
\includegraphics[scale=1,width=8cm]{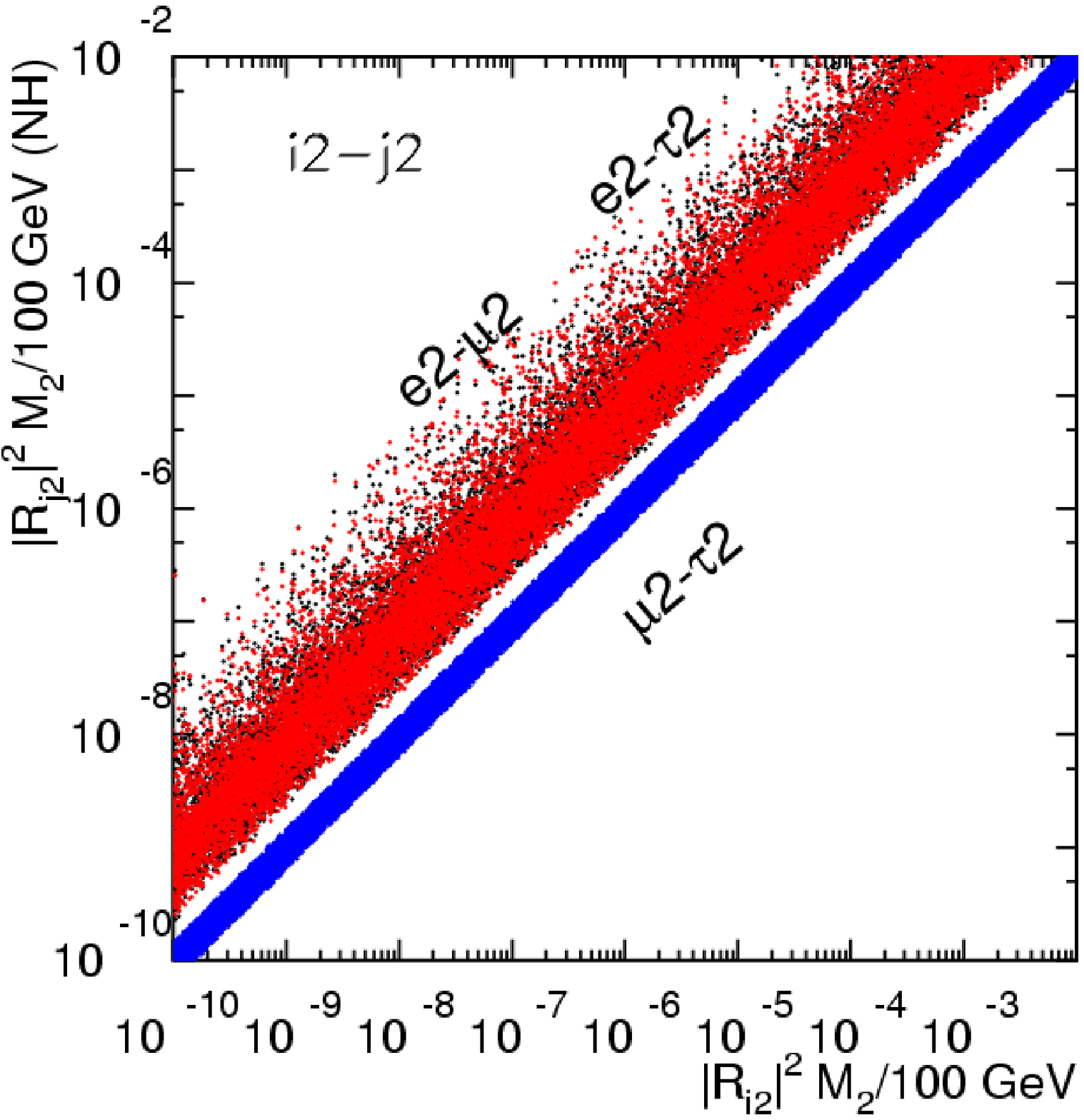}
\includegraphics[scale=1,width=8cm]{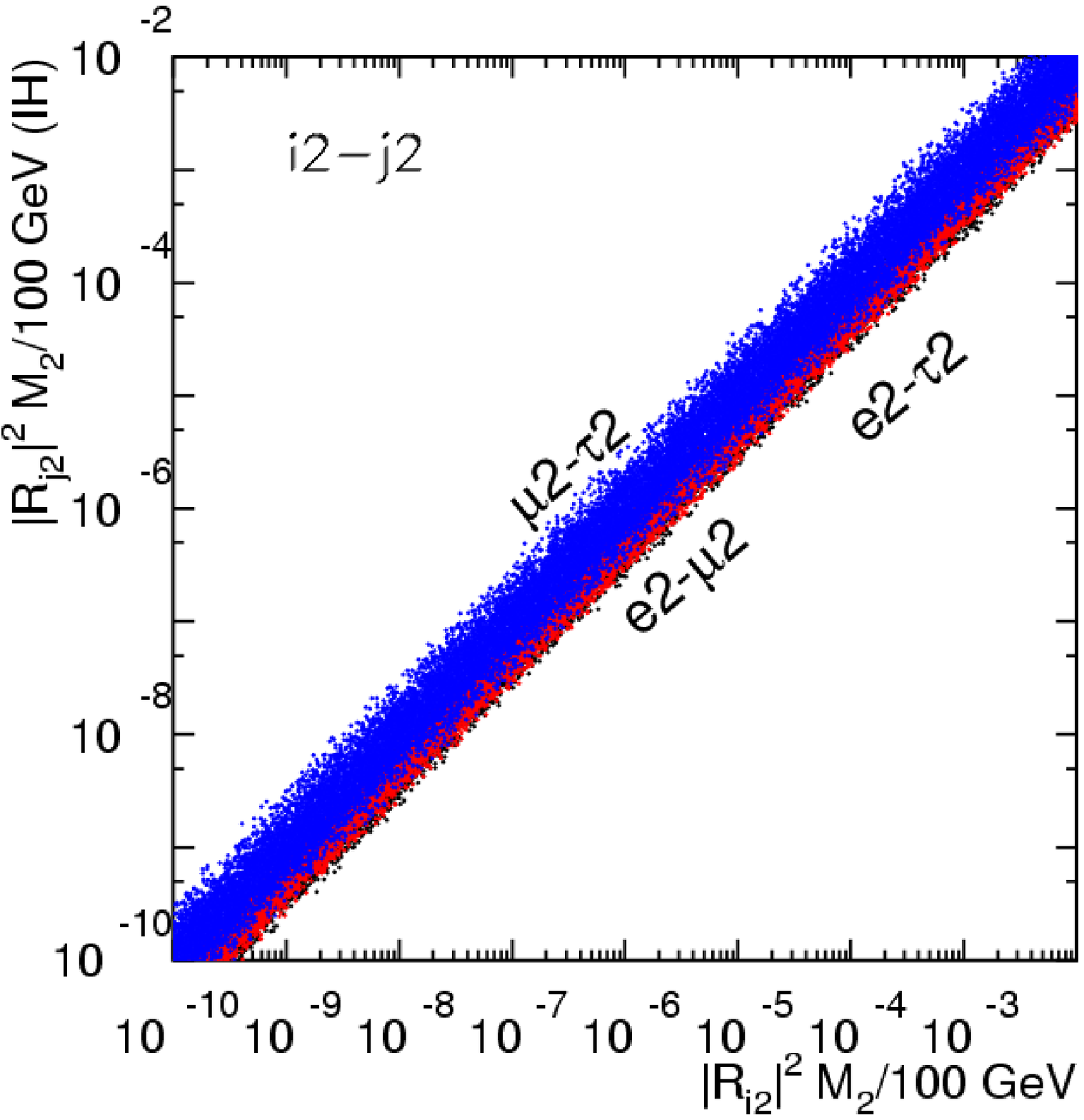}
\end{tabular}
\end{center}
\caption{$|R_{l 2}|^2 M_2/100~{\rm GeV}$ for NH (left) and IH
(right), assuming vanishing Majorana phase and $0\leq x\leq 2\pi,
-20\leq y\leq 20$.} \label{VLlN}
\end{figure}

Using the above experimental constraints, one can explore the
allowed values for the $R_{l i} (l=e,\mu,\tau; i=2,3)$ couplings and
the heavy masses. In Fig.~\ref{VLlN} we plot the allowed values for
the normalized couplings $|R_{l2}|^2M_2/100~{\rm GeV}$ of each
lepton flavor for each spectrum, the NH (left panel) and the IH
(right panel), assuming vanishing Majorana phase. We see that the
couplings of different lepton flavors have strong correlation as
expected.

\begin{figure}[tb]
\begin{center}
\begin{tabular}{cc}
\includegraphics[scale=1,width=8cm]{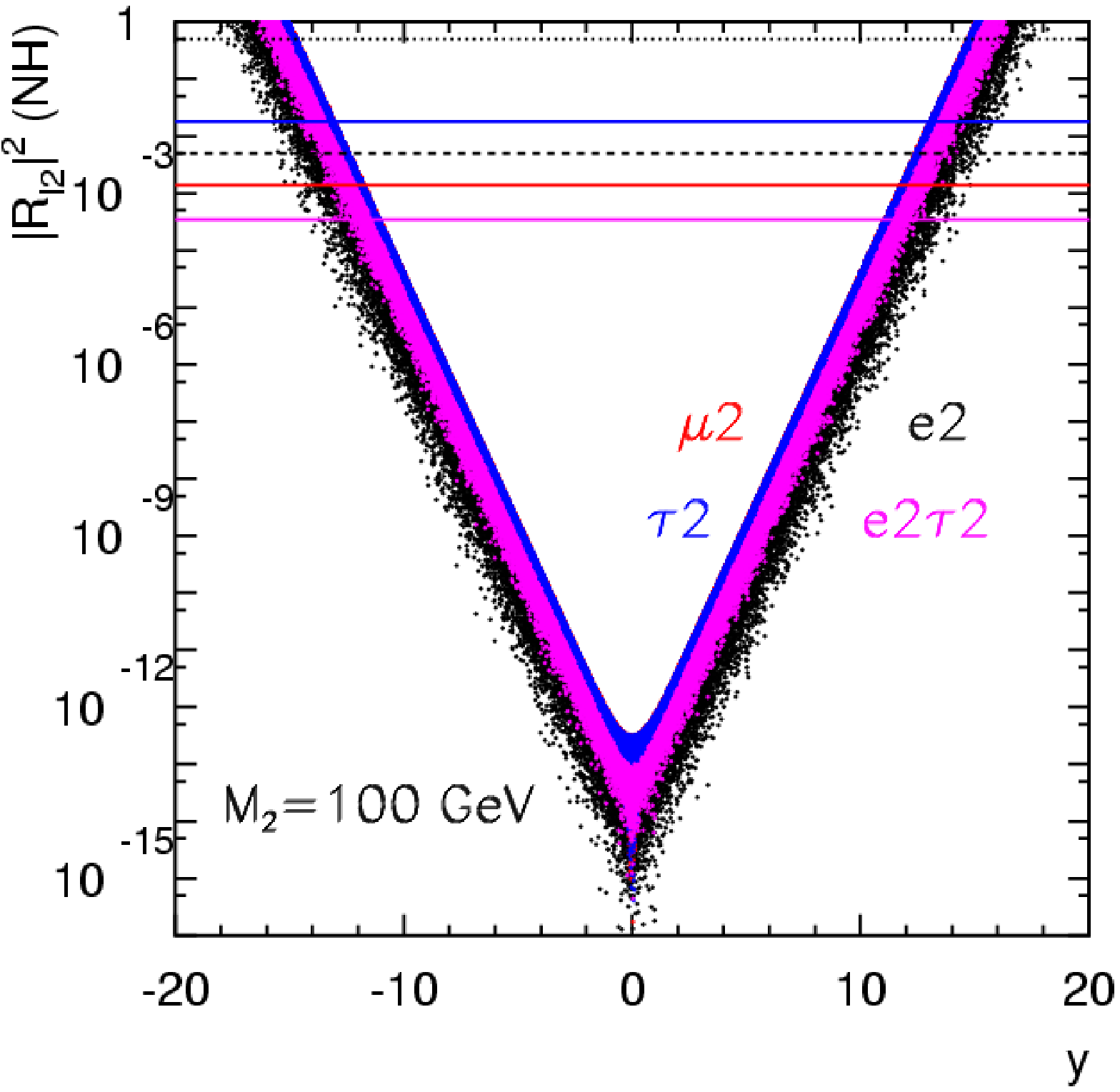}
\includegraphics[scale=1,width=8cm]{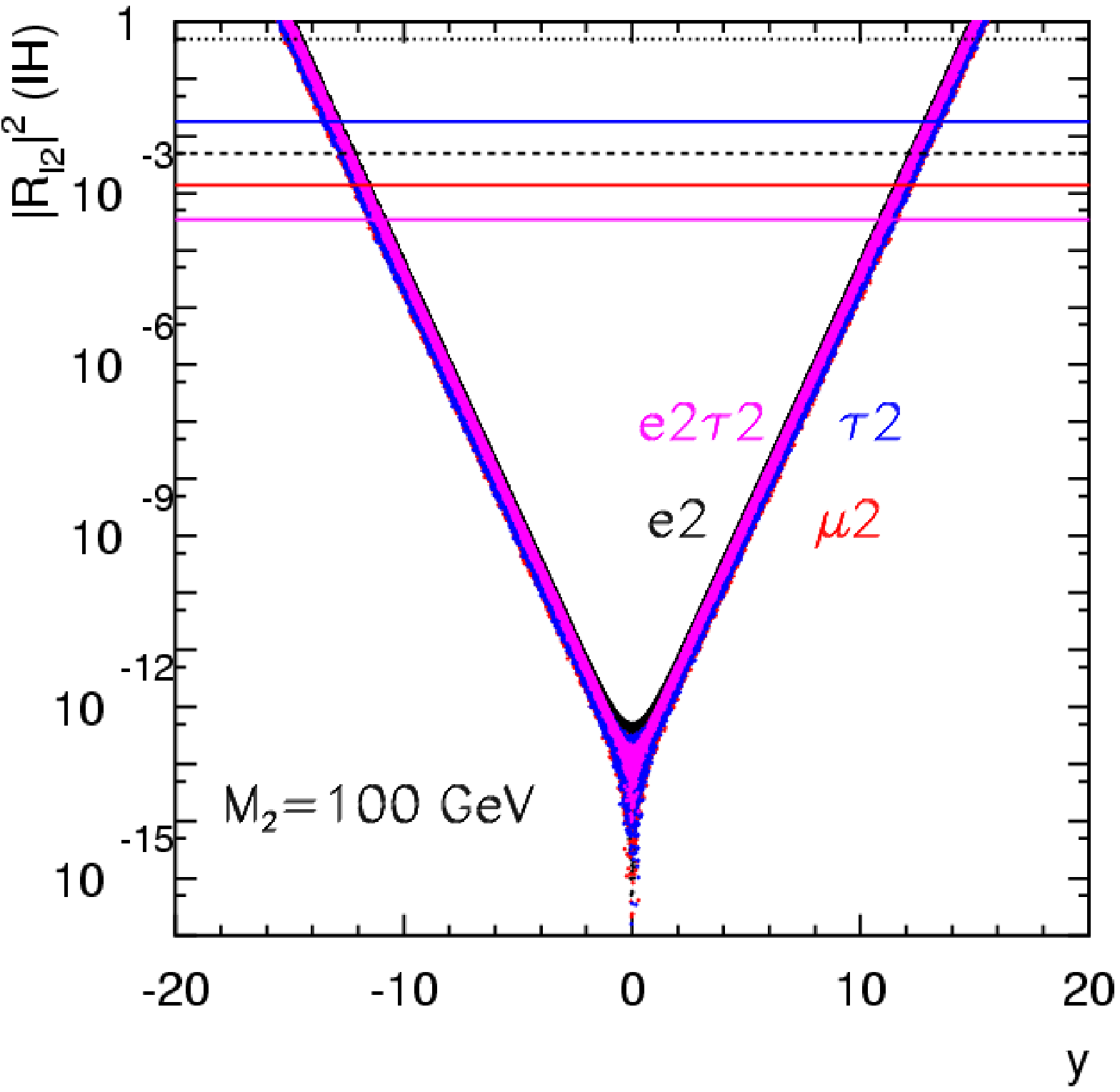}
\end{tabular}
\end{center}
\caption{$|R_{l 2}|^2$ versus parameter $y$ in matrix $O$ for NH
(left) and IH (right), assuming vanishing Majorana phase and $0\leq
x\leq 2\pi, M_2=100~{\rm GeV}$. The horizontal lines represent the
current experimental bounds from precision electro-weak
measurements~\cite{ew} and searches of neutrinoless double
beta-decay, the dashed and dotted lines, assuming $\Delta
M^2_{32}/M^2_2$ to be $10^{-3}$ and $10^{-5}$, respectively .}
\label{VLlNv}
\end{figure}

Moreover, as mixing elements of matrix $R$ govern processes
involving right-handed neutrinos, most stringent model-independent
data from precision electro-weak measurements and low-energy
lepton-number violating processes can constrain unknown parameters
in the complex orthogonal matrix $O$ and the Majorana phase $\Phi$.
We show $|R_{l2}|^2$ as a function of the imaginary part $y$ in
parameter $\theta$ of $O$ in Fig.~\ref{VLlNv}. One can see that
$|R_{l2}|^2$ can reach $\sim 10^{-3}$, the upper bound from
precision measurement. Constraint from neutrinoless double beta
decay is weaker for quasi-degenerate $\nu_{R2,3}$. As noted
previously, heavy neutrinos with $|R_{li}|^2\sim 10^{-3}$ can be
produced and studied at the LHC.

\begin{figure}[tb]
\begin{center}
\begin{tabular}{cc}
\includegraphics[scale=1,width=8cm]{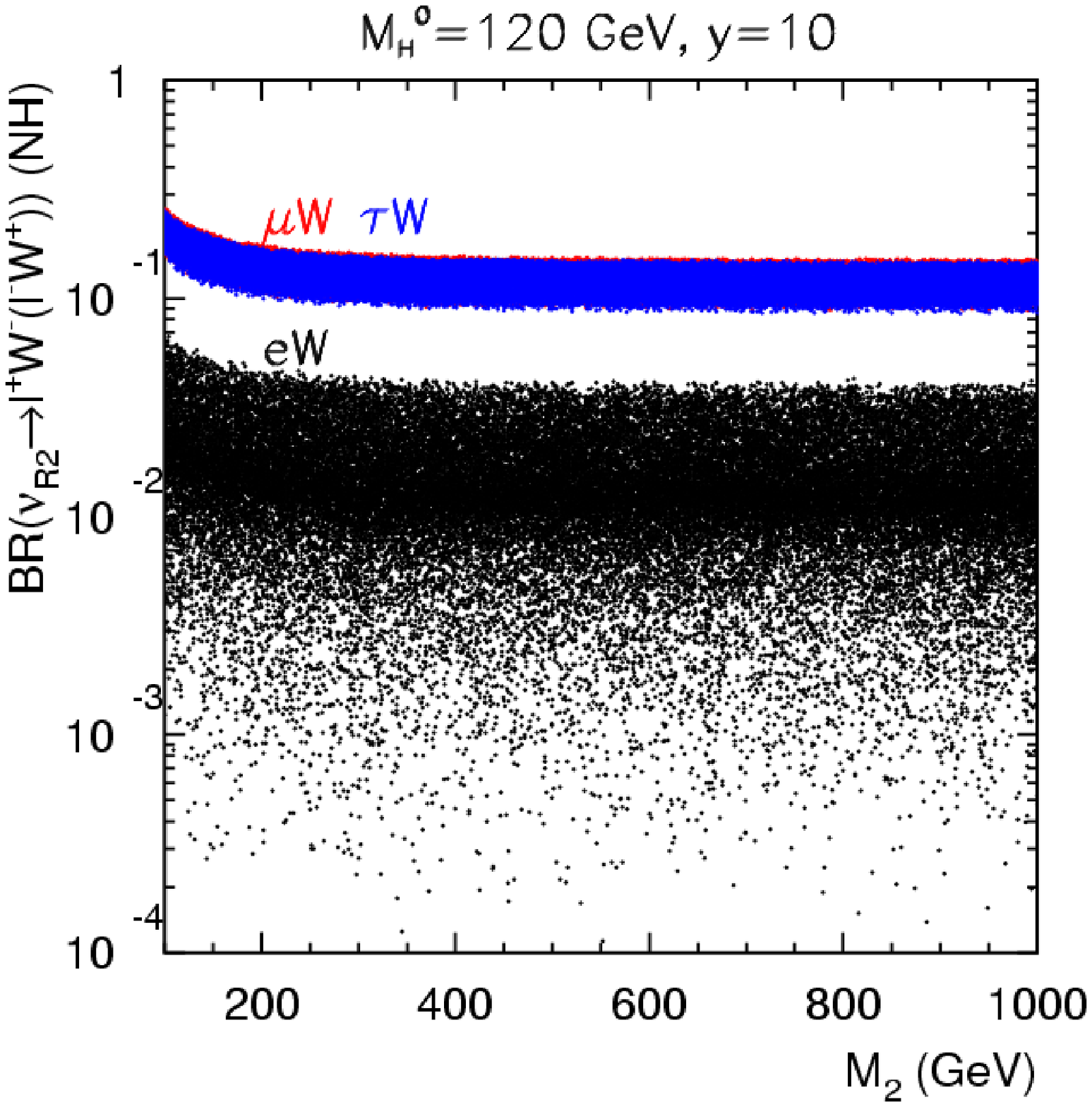}
\includegraphics[scale=1,width=8cm]{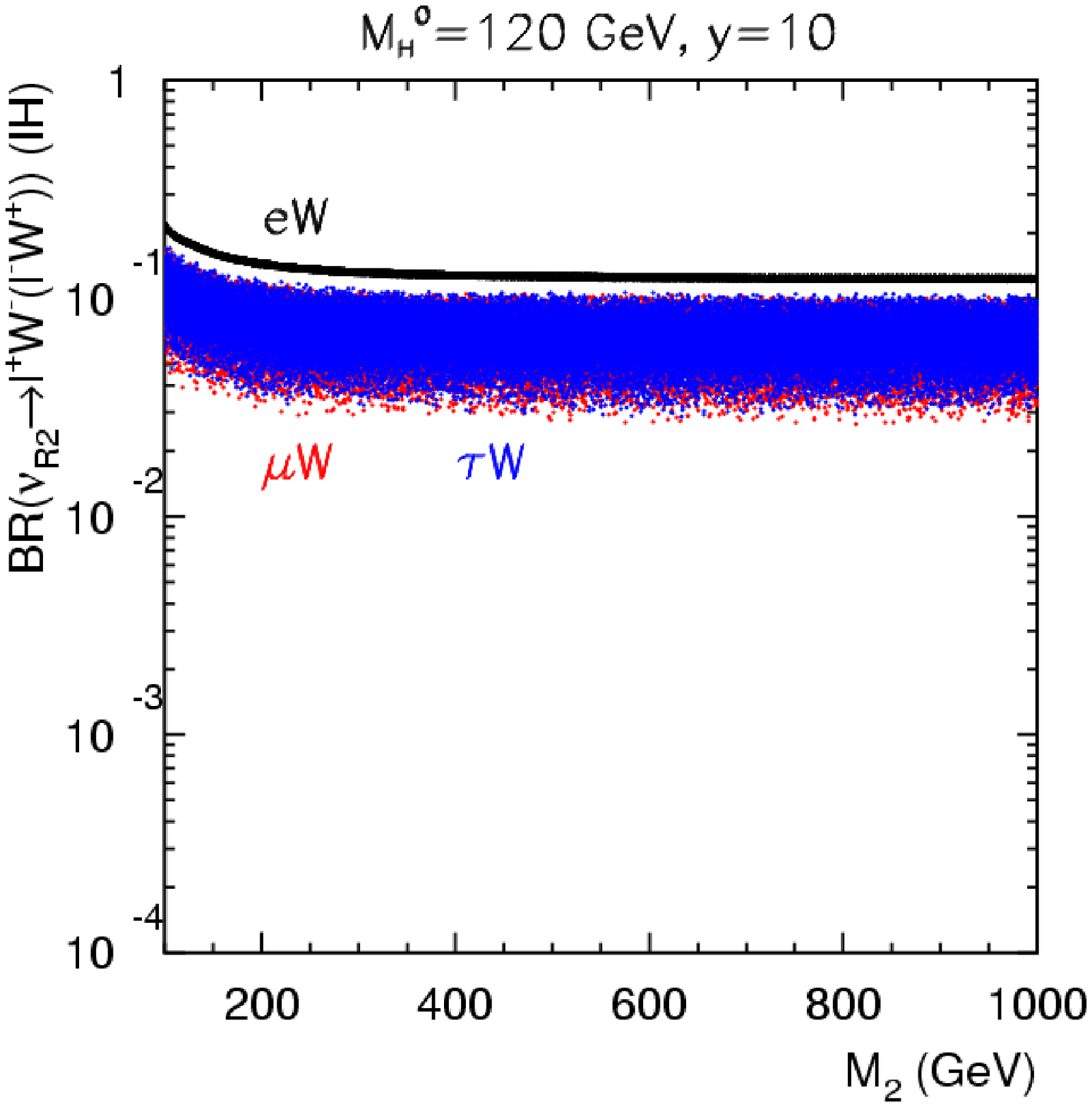}
\end{tabular}
\end{center}
\caption{Branching fractions of degenerate neutrinos $\nu_{R2} \to
l^+ W^-(l^-W^+) \ (l=e,\mu,\tau)$ for NH (left) and IH (right)
versus heavy neutrino mass with $0\leq x\leq 2\pi, y=10$ and
$M_{H}=120~{\rm GeV}$, assuming vanishing Majorana phase.}
\label{nbr}
\end{figure}

\begin{figure}[tb]
\begin{center}
\begin{tabular}{cc}
\includegraphics[scale=1,width=8cm]{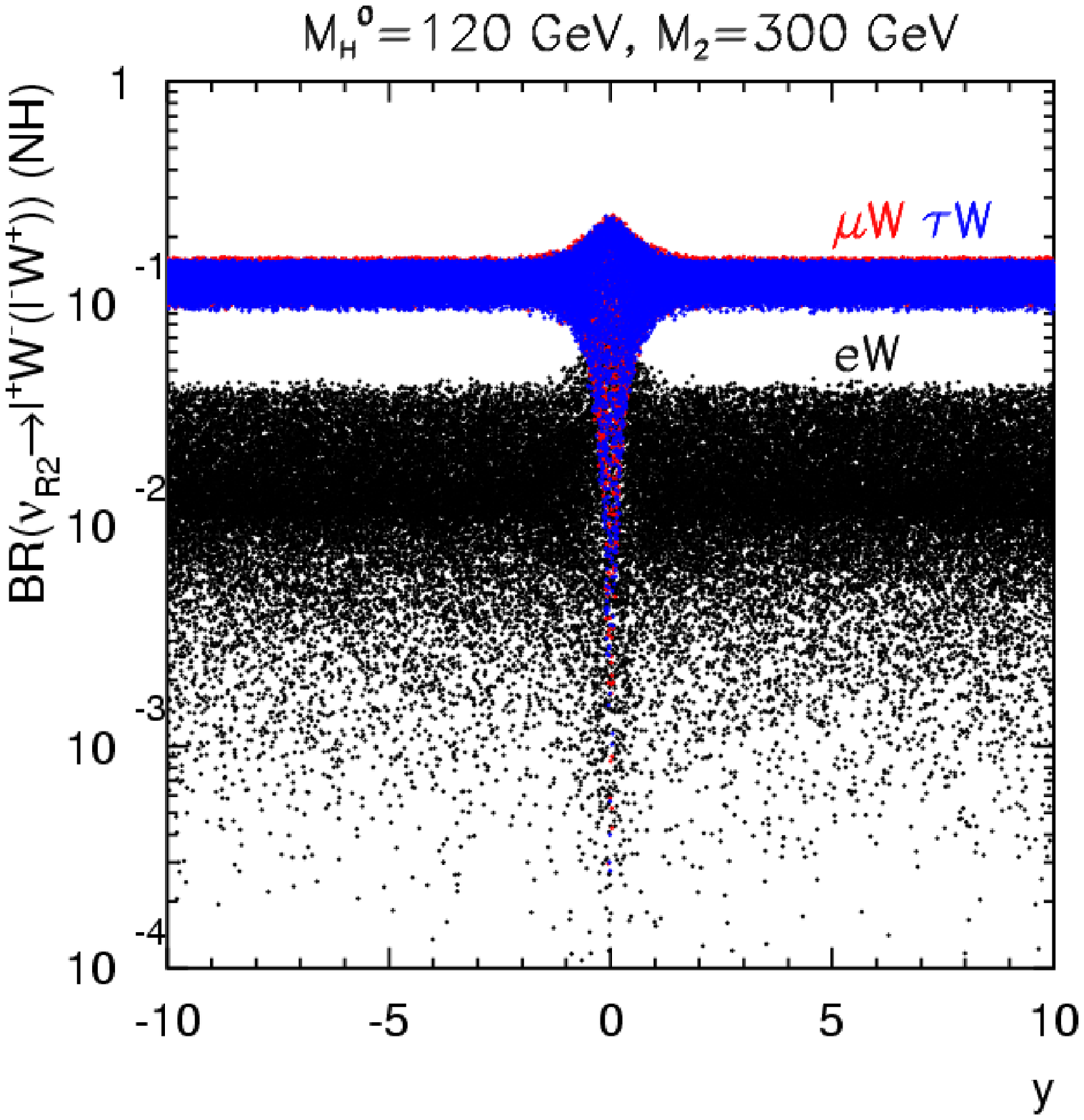}
\includegraphics[scale=1,width=8cm]{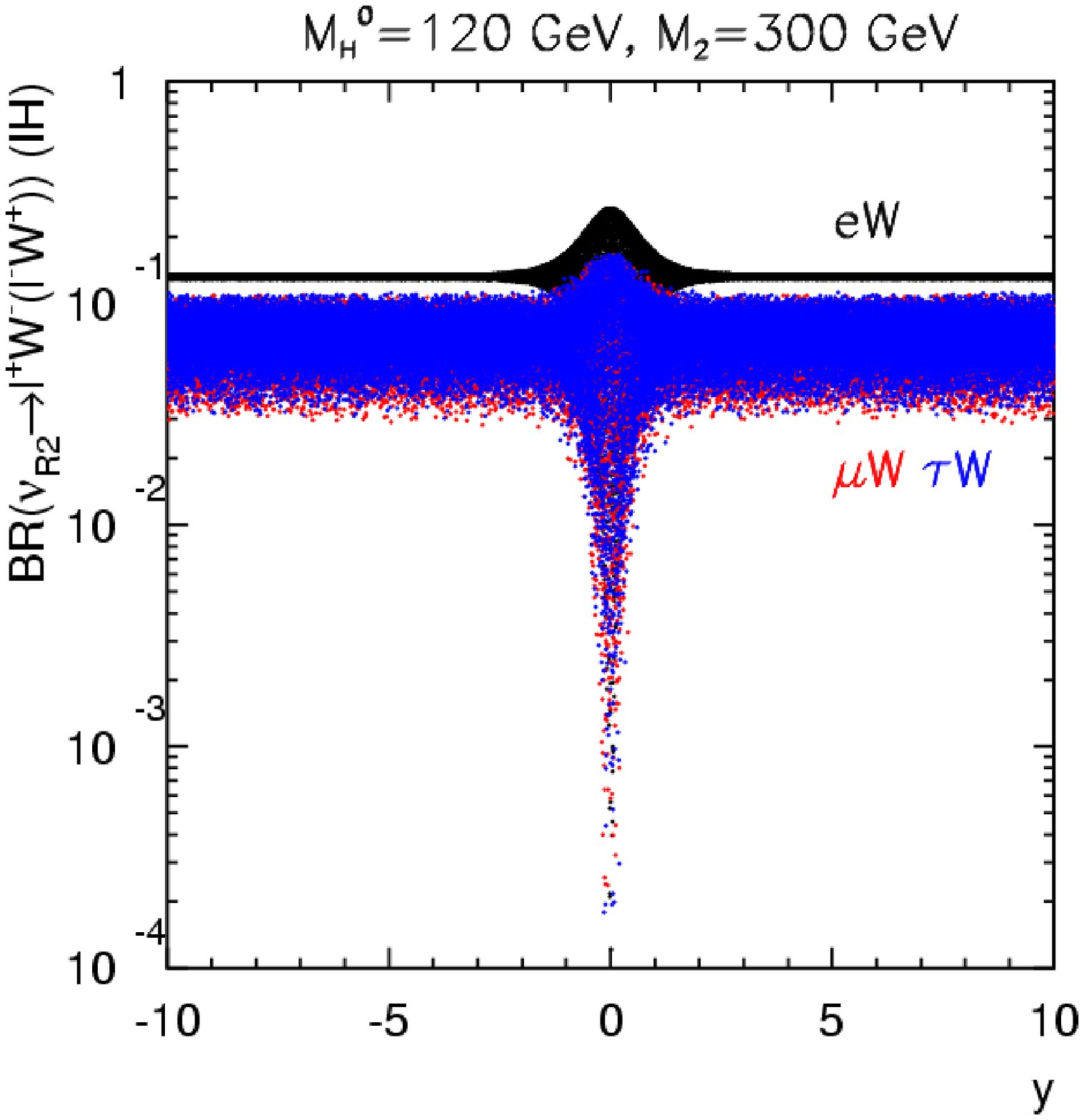}
\end{tabular}
\end{center}
\caption{Branching fractions of degenerate neutrinos $\nu_{Ri} \to
l^+ W^-(l^-W^+) \ (l=e,\mu,\tau)$ for NH (left) and IH (right)
versus parameter $y$ with $0\leq x\leq 2\pi, M_2=300~{\rm GeV}$ and
$M_{H}=120~{\rm GeV}$, assuming vanishing Majorana phase.}
\label{nbrv}
\end{figure}

The partial decay widths of the heavy Majorana neutrinos $\nu_{Ri}$
are given by~\cite{Tao}
 \begin{eqnarray}
\Gamma^{l W_L}&\equiv& \Gamma(\nu_{Ri} \to l^-W_L^+)=
\Gamma(\nu_{Ri} \to l^+W_L^-)= {g^2\over 64\pi M_W^2}|R_{l
i}|^2M_i^3(1-\mu_{iW})^2,
\\
\Gamma^{l W_T}&\equiv&\Gamma(\nu_{Ri} \to l^-W_T^+)={g^2\over
32\pi}|R_{l i}|^2M_i(1-\mu_{iW})^2,
\\
\Gamma^{\nu_l Z_L}&\equiv&\Gamma(\nu_{Ri} \to \nu_l Z_L)={g^2\over
64\pi M_W^2}|R_{l i}|^2M_i^3(1-\mu_{iZ})^2,
\\
\Gamma^{\nu_l Z_T}&\equiv&\Gamma(\nu_{Ri} \to \nu_l Z_T)={g^2\over
32\pi c_W^2}|R_{l i}|^2M_i(1-\mu_{iZ})^2,
 \end{eqnarray}
where $\mu_{ij}=M_j^2/M_i^2$. If $\nu_{Ri}$ is heavier than the
Higgs boson $H^0$, one has the additional channels
\begin{eqnarray}
\Gamma^{\nu_l h}&\equiv& \Gamma(\nu_{Ri} \to \nu_l H^0)={g^2\over
64\pi M_W^2}|R_{l i}|^2M_i^3(1-\mu_{iH})^2.
\end{eqnarray}
As discussed above, the lepton-flavor contents of $\nu_{R}$ decays
will be different in each neutrino spectrum. In order to search for
the events with best reconstruction, we will only consider the
$\nu_R$ decay to charged leptons plus a $W^\pm$. In Fig.~\ref{nbr}
we show the impact of the neutrino masses and mixing angles and
parameters in $O$ on the branching fractions of the heavy neutrino
$\nu_{R2}$ decaying into $e,\mu,\tau$ lepton plus $W$ boson,
respectively, with the left panels for the NH and the right panels
of the IH, assuming vanishing Majorana phase. We also plot the
dependence of branching fractions on parameter $y$ in
Fig.~\ref{nbrv}. For large values of heavy neutrino mass or $y$ the
branching fractions can differ by one order of magnitude in NH case
$BR(\mu^\pm W^\mp),BR(\tau^\pm W^\mp)\gg BR(e^\pm W^\mp)$ and about
a factor of few in the IH spectrum $BR(e^\pm W^\mp)>BR(\mu^\pm
W^\mp),BR(\tau^\pm W^\mp)$.

\begin{figure}[tb]
\begin{center}
\begin{tabular}{cc}
\includegraphics[scale=1,width=8cm]{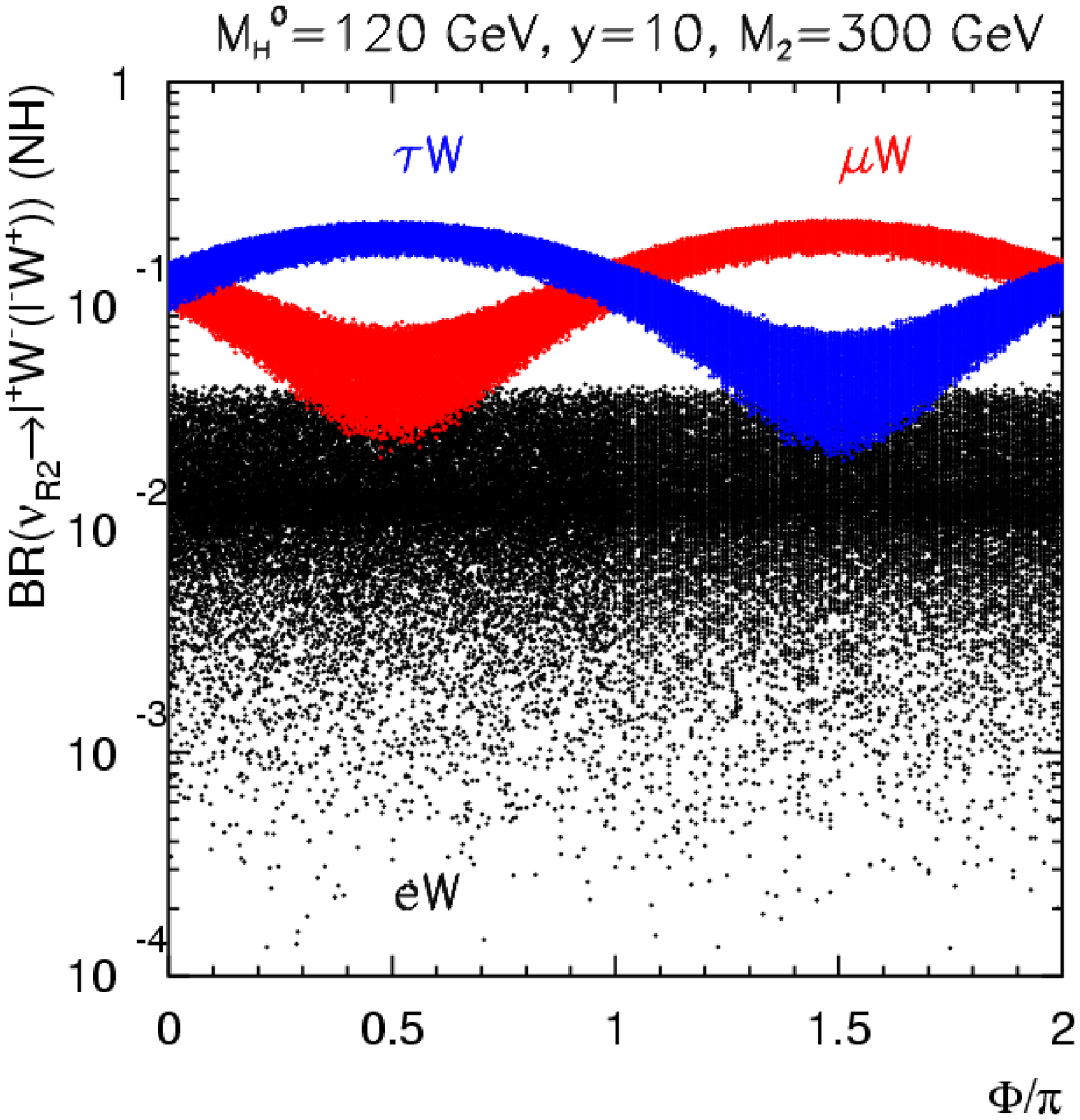}
\includegraphics[scale=1,width=8cm]{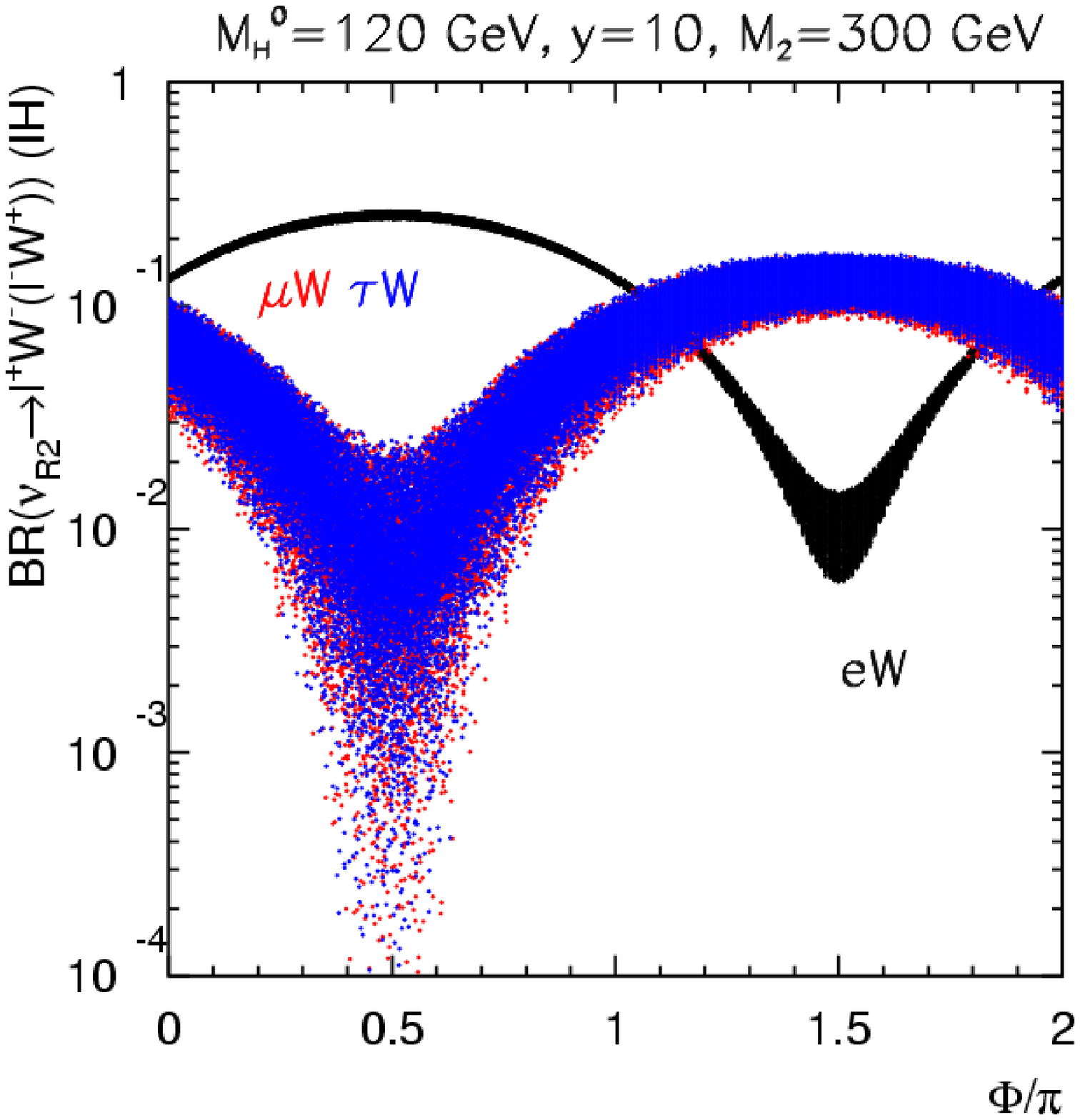}
\end{tabular}
\end{center}
\caption{Branching fractions of $\nu_{R2} \to l^+ W^-(l^-W^+) \
(l=e,\mu,\tau)$ versus Majorana phase $\Phi$ for NH (left) and IH
(right) when $M_2=300~{\rm GeV}, M_H=120~{\rm GeV}$ and $0\leq x\leq
2\pi, y=10$.} \label{nbrph}
\end{figure}

In general, the $\nu_{R}$ decay rates depend on only one Majorana
phase $\Phi$ when the lightest neutrino mass vanishes in the NH or
IH case. In Fig.~\ref{nbrph}, we show the dependence of $\nu_{R2}$
decay branching fractions on Majorana phase $\Phi$ in NH and IH for
$y=10$. In NH the dominant channels swap from $\tau^\pm W^\mp$ when
$\Phi\approx \pi/2$ to $\mu^\pm W^\mp$ when $\Phi\approx 3\pi/2$ by
a few times. In IH the dominant channels swap from $e^\pm W^\mp$
when $\Phi\approx \pi/2$ to $\mu^\pm W^\mp,\tau^\pm W^\mp$ when
$\Phi\approx 3\pi/2$ by more than one order time of magnitude.
Moreover, it is important to note that the curves of branching
fractions corresponding to Majorana phase translate parallelly by a
phase $\pi$ for $-y$ case. This qualitative change can be made use
of extracting the value of the Majorana phase $\Phi$ and parameter
$y$.

If nature indeed uses low scale heavy neutrinos, of order one
hundred GeV, with large mixing to light neutrinos, they may be
produced at the LHC. The $\nu$SM can be tested by studying
correlations between the decay patterns of heavy neutrinos and light
neutrino mass hierarchies.

\vspace{0.3 cm}
 \noindent {\bf Conclusions}

 In summary we have studied the symmetry and some phenomenologies
 of the $\nu$SM.  The minimal model of this type is a 3 + 3 seesaw model with
 one of the right-handed neutrino to be light to play the role of the keV warm DM.
 We found that the mass and mixing parameters of the
 DM right-handed neutrino, constrained by experimental data, lie in the range
 that there is an approximate FL symmetry in the Lagrangian
 of the $\nu$SM. The masses of the light active neutrino mass hierarchy, predicted
 in $\nu$SM, can be IH or NH, but can not be quasi-degenerate.
 The seesaw masses of the light active neutrinos
 are dominated by contributions of two multi-GeV right-handed
 neutrinos $\nu_{R_{2,3}}$ and can be approximated by $3+2$ seesaw formula.
  We should emphasize that the approximate FL symmetry discussed in the
 present work is not only valid for the range of $\nu_{R2,3}$ masses discussed,
 it is also valid for seesaw model with much larger or smaller masses of $\nu_{R_{2,3}}$,
 as long as the keV scale dark matter constrained by observational data
 is included in the seesaw model.
 The keV scale $\nu_{R1}$ has a number of astrophysical
 implications, such as effect on supernova explosion~\cite{supernova}
 and on the re-ionization~\cite{reionization}.
 The phenomenology of the two multi-GeV scale right-handed
 neutrinos is similar to the phenomenology of $3+2$ seesaw model.
 The lightest light neutrino has a mass of order $10^{-5}$ eV.
 We found that in $\nu$SM the Yukawa couplings can be large.
 In particular we found that double beta decay experiment does not
 give strong constraint on the mixings and masses of two multi-GeV
 scale right-handed neutrinos if they are degenerate or quasi-degenerate. The
 Yukawa couplings can be large enough to be tested in LHC experiments.
 We also found that there are
 strong correlations between the couplings of the two heavy neutrinos
 $\nu_{R_{2,3}}$ and light neutrino mass hierarchy.
 The decay patterns of these heavy neutrinos sensitively depend on the Majorana phase.
 The decay patterns of the
 right-handed neutrinos $\nu_{R_{2,3}}$ can be used to extract information
 of the mass hierarchy and Majorana phase of the light active
 neutrinos.

\acknowledgments
 This work is supported in part by NSC and NCTS, the Science and Technology
 Commission of Shanghai Municipality under contract number 09PJ1403800 and
 National Science Foundation of China(NSFC), grant 10975052.\\

Note added:
 After the submission of the present article, recent analysis on X-ray
 observations of local dwarf Willian 1 show evidence that the
 $\nu_{R_1}$ dark matter may have mass around $5$ keV with mixing
 $|R_{l1}|^2\sim 10^{-9}$ ~\cite{Xray1}. Another analysis on the
 X-ray observation of the galactic center suggests that $\nu_{R1}$
 dark matter has mass around $17$ keV with mixing $|R_{l1}|^2\sim
 10^{-12}$~\cite{Xray2}. It's easy to see that Eq. (\ref{constraint1}) is valid
 for these two groups of parameters of $\nu_{R_1}$ and the approximate Friedberg-Lee symmetry
 discussed in this article applies to models
 with these ranges of parameters space.


\end{document}